\documentclass[prd, twocolumn, superscriptaddress,floatfix, nofootinbib, preprintnumbers]{revtex4-2}
\pdfoutput=1
\usepackage[utf8]{inputenc}
\usepackage{hyperref}
\hypersetup{colorlinks=true,citecolor=blue}\usepackage{rotating}
\usepackage{array}
\usepackage{amsmath,amssymb,amstext}
\usepackage[normalem]{ulem}
\usepackage{slashed}
\usepackage{booktabs}
\usepackage[pdftex,table,dvipsnames]{xcolor}
\usepackage{multirow}
\usepackage{url}
\usepackage{color}
\usepackage{xspace}
\usepackage{comment}
\usepackage{enumitem}
\usepackage[caption=false]{subfig}
\usepackage{siunitx}
\usepackage{graphicx}
\usepackage{placeins}
\usepackage{layouts}

\usepackage{multirow}
\begin{document}


\title{How to pick the best anomaly detector?}

\author{Marie Hein}
\email{marie.hein@rwth-aachen.de}
\affiliation{Institute for Theoretical Particle Physics and Cosmology, RWTH Aachen University, D-52056 Aachen, Germany}

\author{Gregor Kasieczka}
\email{gregor.kasieczka@uni-hamburg.de}
\affiliation{Institut f\"{u}r Experimentalphysik, Universit\"{a}t Hamburg, 22761 Hamburg, Germany}

\author{Michael Kr\"amer}
\email{mkraemer@physik.rwth-aachen.de}
\affiliation{Institute for Theoretical Particle Physics and Cosmology, RWTH Aachen University, D-52056 Aachen, Germany}

\author{Louis Moureaux}
\email{louis.moureaux@cern.ch}
\affiliation{Institut f\"{u}r Experimentalphysik, Universit\"{a}t Hamburg, 22761 Hamburg, Germany}

\author{Alexander M\"uck} 
\email{mueck@physik.rwth-aachen.de}
\affiliation{Institute for Theoretical Particle Physics and Cosmology, RWTH Aachen University, D-52056 Aachen, Germany}

\author{David Shih}
\email{shih@physics.rutgers.edu}
\affiliation{NHETC, Dept.\ of Physics and Astronomy, Rutgers University, Piscataway, NJ 08854, USA}

\begin{abstract}
Anomaly detection has the potential to discover new physics in unexplored regions of the data. However, choosing the best anomaly detector for a given data set in a model-agnostic way is an important challenge 
which has hitherto largely been neglected. In this paper, we introduce the data-driven ARGOS metric, which has a sound theoretical foundation and is empirically shown to robustly select the most sensitive anomaly detection model given the data. Focusing on weakly-supervised, classifier-based anomaly detection methods, we show that the ARGOS metric outperforms other model selection metrics previously used in the literature, in particular the binary cross-entropy loss. We explore several realistic applications, including hyperparameter tuning as well as architecture and feature selection, and in all cases we demonstrate that ARGOS is robust to the noisy conditions of anomaly detection.
\end{abstract}

\maketitle

\section{Introduction}
\label{sec:Introduction}

In recent years there has been an explosion of new methods for model-agnostic new physics searches at the LHC, leveraging powerful modern machine-learning (ML) anomaly detection methods such as autoencoders and weak supervision \cite{Kasieczka:2021xcg, Aarrestad:2021oeb, Belis:2023mqs, hepmllivingreview}.
These methods are able to find anomalies in the data with varying degrees of model independence: either signal-model-independence, in that they make no use of any specific new physics models in their training, and/or background-model-independence in that they make no use of Standard Model (SM) simulations in their training. As such, these new model-agnostic search strategies offer significant new opportunities for discovery in the vast, unexplored phase space of the LHC data, beyond what model-specific search strategies have been able to achieve to date. 

However, one important question has been left largely unaddressed to date: how does one pick the best anomaly detector from all the available options? In our case, we consider the ``best" anomaly detector to be the {\it most sensitive to unknown signals}, as measured by the standard Significance Improvement Characteristic (SIC):
\begin{equation}
\text{SIC}=\frac{\epsilon_S}{\sqrt{\epsilon_B}},
\end{equation}
where $\epsilon_{S/B}$ is the signal/background efficiency after selecting the most anomalous events with the anomaly detector. The SIC is a multiplicative factor that quantifies the extent to which the anomaly detector enhances the significance of the new physics hiding in the data. Our goal is to select an anomaly detector that can achieve a high SIC sensitivity to any signal that is actually present in the given dataset.

So far, when it comes to different approaches (e.g.\ autoencoders vs.\ weak supervision), the idea has been to just try them all, as seen in Ref.~\cite{CMS:2024nsz}.
Within a given approach, there are myriad smaller but still vital choices to make: which model epoch to pick during training, which model architecture and hyperparameters to use, or even which features to use.
Metrics traditionally used to guide these c qhoices, such as the SIC and area under curve (AUC), are based on truth labels and require the use of benchmark signals.
This is clearly not satisfactory for anomaly detection as the added signal dependence may, in the worst case, work against finding a signal actually present in the data.
 
Consequently, none of the existing experimental anomaly detection analyses~\cite{ATLAS:2020iwa,ATLAS:2023azi,ATLAS:2023ixc,CMS:2024nsz,ATLAS:2025obc,ATLAS:2025kuz,CMS:2025dbd,CMS:2025eou} report systematic model optimization.
Model architectures and training parameters are generally derived from the original publication introducing the corresponding method, with adjustments to network capacity to deal with the larger experimental datasets. In some cases~\cite{ATLAS:2023ixc,CMS:2024nsz,CMS:2025dbd}, parameters have been retuned using a small set of benchmark signals. In Refs.~\cite{CMS:2024nsz,CMS:2025dbd}, anomaly scores have been selected from different classifier runs or epochs in certain approaches -- CWoLa Hunting and TNT used the number of events selected in the signal region at a fixed background efficiency in the side bands as a metric; and CATHODE  used the validation loss, following Ref.~\cite{Hallin:2021wme}. In Ref.~\cite{Grosso:2024wjt}, combining different anomaly detection models has been studied in the context of the NLPM methodology~\cite{DAgnolo:2018cun}.

In this paper we introduce ARGOS (Above Random Gain Of SIC), a new {\it fully data-driven} metric for selecting the best anomaly detector. 
The key ingredient (besides the anomaly detector) that is required to calculate ARGOS is what we refer to as a background template (BT), which is a sample of events following the distribution of SM background in the signal region (SR). 
Then ARGOS is defined to be:
\begin{equation}\label{eq:ARGOS}
    \text{ARGOS} = \frac{\epsilon_{\text{SR}}}{\sqrt{\epsilon_{\text{BT}}}}-\sqrt{\epsilon_{\text{BT}}},
\end{equation}
where $\epsilon_{\text{SR/BT}}$ is the efficiency to select SR/BT events for a given anomaly threshold. By definition, ARGOS is equivalent to the standard test statistic for an excess in a cut-and-count experiment with $N_{\text{SR}}$ signal region events, where the observed number of events $\epsilon_{\text{SR}} N_{\text{SR}}$ after the anomaly cut is compared to the expected event number $\epsilon_{\text{BT}} N_{\text{SR}}$ for the null hypothesis. This test statistic has also been used for specific anomaly detection methods~\cite{Das:2024eie,Grosso:2025kmt}. Here, however, we advocate ARGOS as a metric for optimizing anomaly detection performance. In contrast to SIC, which cannot be calculated for real (unlabeled) data, ARGOS is accessible for any anomaly detector if there is a corresponding background template in addition to the (unlabeled) data set. In Section~\ref{sec: optimization metrics}, we show that ARGOS has a sound theoretical foundation, being monotonic with the actual signal-to-background SIC for any unknown signal at a fixed background efficiency when the background template is perfect. More generally, ARGOS is proportional to SIC to a very good approximation in any meaningful setting.

For concreteness, we demonstrate the utility of ARGOS within the context of weak supervision~\cite{Metodiev:2017vrx}, where a classifier is trained to distinguish two mixed-label data sets, for resonant anomaly detection. However, nothing in principle limits ARGOS to this context. Using the LHCO dataset~\cite{Kasieczka:2021xcg}, introduced in Section~\ref{sec:setup}, we demonstrate how ARGOS can consistently select the best hyperparameters, architecture (e.g.\ neural networks vs.\ different boosted decision trees) and model epoch in Section~\ref{sec:tests}. As an outlook, we also briefly discuss the potential for ARGOS to select the best features in a fully data-driven way before we conclude in Section~\ref{sec:Conclusion}. Technical aspects of our experiments are discussed in several Appendices.

\section{The ARGOS Metric}\label{sec: optimization metrics}

In this Section, we show that ARGOS, as introduced in Sec.~\ref{sec:Introduction}, is the optimal data-driven metric for evaluating anomaly detectors when using an ideal BT. For this purpose, we rewrite the ARGOS metric as given in Eq.~(\ref{eq:ARGOS}) in terms of the background and signal efficiencies using the probability density of the SR 
\begin{align}\label{eq:datasets weak supervision}
    p_{\text{SR}}(x) &= f_S\cdot p_S(x)+ (1-f_S)\cdot p_B(x) \, ,
\end{align}
where $f_S$ is the signal fraction in the signal region and $p_{S/B}(x)$ denote the probability density for the signal/background events. For an ideal BT, we have
\begin{align}\label{eq:idealized}
p^\text{ideal}_{\text{BT}}(x)&=p_B(x) \, ,
\end{align}
which we try to approximate in real-world applications as discussed in Sec.~\ref{sec:setup}. Using these densities, which imply $\epsilon_{\text{BT}}=\epsilon_B$, we find
\begin{align} 
    \text{ARGOS}&=\frac{f_S\cdot\epsilon_{S}+(1-f_S)\cdot\epsilon_B}{\sqrt{\epsilon_{B}}}-\sqrt{\epsilon_{B}}\\ \label{eq:ARGOSandSIC}
    &= f_S\cdot \left(\text{SIC}-\sqrt{\epsilon_B}\right),
\end{align}
Thus we see that ARGOS and SIC are closely related, at least in the idealized setting.

In more detail, there are multiple ways one can view this result:
\begin{itemize}
\item At fixed $\epsilon_B$, ARGOS is monotonic with SIC, since the signal fraction $f_S$ is a constant. Then the best anomaly detector with respect to the ARGOS metric is also the best anomaly detector with respect to SIC. Additionally, ARGOS and the number of selected signal region events are also monotonic in this case and therefore also result in the same optimal decisions. The latter was used in Ref.~\cite{CMS:2024nsz, CMS:2025dbd} to select epochs from multiple trainings of the CWoLa Hunting and TNT approaches.

\item  For ARGOS, however, there is no need to fix the background efficiency a priori. For a good anomaly detector, $\sqrt{\epsilon_B}\ll {\rm SIC}$ also holds, such that the maximum value of ARGOS as a function of score threshold is also close to the maximum value of SIC. Then an added benefit of using the ARGOS metric in this way is that one can also select the optimal working point of the anomaly detector in a data-driven way. In Ref.~\cite{Grosso:2025kmt}, instead anomaly detectors with different background efficiencies have been combined using the corresponding test statistic.

\end{itemize}

In Section~\ref{sec:tests}, we take the second approach and employ the maximum value of ARGOS for optimization. We will show that it retains good performance for realistic BTs.

\section{Setup}\label{sec:setup}

\subsection{Data set}\label{subsec:dataset}
In Section \ref{sec:tests}, we perform experiments in a number of different scenarios, which test different metrics for optimizing the setup for anomaly detection. 
These experiments are performed using the LHC Olympics 2020 (LHCO) \cite{Kasieczka:2021xcg} R\&D data set \cite{LHCOdataset}, which is a common benchmark data set for LHC anomaly detection. It contains $10^6$ QCD dijet background events and $10^5$ signal events. The signal is a $W'$ resonance at $m_{W'} = 3.5\,\text{TeV}$ with the decay $W' \rightarrow X(\rightarrow qq)Y(\rightarrow qq)$, where $X$ and $Y$ have masses $m_{X}=100\,\text{GeV}$ and $m_{Y}=500\,\text{GeV}$, respectively. Additionally, we use {612858} SR background events from Ref.~\cite{extraLHCOdataset}, which were generated for Ref.~\cite{Hallin:2021wme}.

The simulation pipeline consists of \texttt{Pythia 8}~\cite{Sjostrand:2006za,Sjostrand:2007gs} and \texttt{Delphes 3.4.1}~\cite{deFavereau:2013fsa}. Using \texttt{Fastjet}~\cite{Cacciari:2011ma}, reconstructed particles are clustered with the anti-$k_{T}$ algorithm~\cite{Cacciari:2005hq} with a distance parameter of $R=1$. All events must pass a leading-jet trigger using  $p_{\mathrm{T}}>\SI{1.2}{\tera\electronvolt}$. For the classification, we use the features also used in Ref.~\cite{Hallin:2021wme}. We select the two highest $p_T$ jets and use the mass of the lighter jet $m_{J_1}$, the mass differences of the two jets $\Delta m_J=m_{J_2}-m_{J_1}$ as well as both jets' 21-subjettiness ratios \cite{Thaler:2010tr, Thaler:2011gf}. In Sec.~\ref{sec:feature selection}, additional feature sets based on Ref.~\cite{Finke:2023ltw} are used. The dijet mass $m_{JJ}$ is used as the resonant feature. 

For the anomaly detection scenarios, we use all 1M background events and inject $N_{sig}$ signal events. We use the same signal events in all runs in analogy to the fixed data set available in an analysis. We define our signal region using the dijet mass by requiring $3.3\,\text{TeV} \le m_{JJ} \le 3.7\,\text{TeV}$, i.e., a $0.4\,\text{TeV}$ window around the $m_{W'}$ resonance, resulting in approximately \num{120000} SR background events. The additional SR background events from Ref.~\cite{extraLHCOdataset} constitute the background template for the IAD (see Sec.~\ref{subsec:IAD}), and also form part of the test set, for which a total of 340\,000 SR background and 20\,000 SR signal events are used.

\subsection{Weakly Supervised Methods for Resonant Anomaly Detection }\label{subsec: weak supervision methods}

We focus on classifier-based resonant anomaly detection in this work, since these methods are predicated on the existence of an accurate BT that describes the smooth background in the SR. For resonant anomalies, the data-driven construction of the BT is usually based on adjacent sideband (SB) regions that facilitate smooth interpolation into the SR. There are a variety of different methods of constructing the background template \cite{Metodiev:2017vrx,Collins:2018epr,Collins:2019jip,Nachman:2020lpy,Andreassen:2020nkr,1815227,Hallin:2021wme,Raine:2022hht,Hallin:2022eoq,Golling:2022nkl,Golling:2023yjq, Das:2024fwo, Leigh:2024chm, Oleksiyuk:2025pmu}, and we use the following three to illustrate the utility of the ARGOS metric:

\subsubsection{Idealized Anomaly Detector}\label{subsec:IAD}
The idealized anomaly detector (IAD) \cite{Hallin:2021wme} is used as a benchmark which is available for simulated data only. 
Instead of constructing the BT in a data-driven way, the IAD simply uses background events simulated with the same pipeline that generated the background data in the SR. Hence, by definition, $p^\text{IAD}_{\text{BT}}(x)=p^\text{ideal}_{\text{BT}}(x)=p_B(x)$. We use approximately \num{272000} such events. 

\subsubsection{CWoLa Hunting}
CWoLa Hunting \cite{Collins:2018epr, Collins:2019jip} directly uses data from short sidebands of $0.2\,\text{TeV}$ on either side of the signal region as the background template, assuming there is little correlation of the features with the dijet invariant mass. Using short sidebands reduces correlations with the dijet mass. As these short sidebands are in total as wide as the SR, the resulting background template has a similar size to the SR.

\subsubsection{CATHODE}
CATHODE \cite{Hallin:2021wme} uses conditional density estimation to interpolate the SB distribution into the SR: as the SB contains only very low amounts of signal events, learning $p(x|m\in SB)$ is largely equivalent to learning the background distribution $p_B(x|m)$, which can then be sampled in the SR to obtain background events in the SR. The density estimator can be used to generate more BT events than there are SR events in order to obtain better background statistics. We use a fixed oversampling factor of four as in Ref.\ \cite{CMS:2024nsz}. The implementation of the density estimation is discussed in App.~\ref{app:implementation}.

\subsection{Classifiers}\label{subsec:classifiers}
We use a total of three different classifier setups for our test cases: 
\begin{enumerate}
    \item A standard multi-layer perceptron (MLP) neural network (NN) with three hidden layers, which is trained using the binary cross-entropy (BCE) loss with the Adam optimizer for 100 epochs. We ensemble the best 10 epochs according to our different metrics, following the ensembling strategy used in Ref.~\cite{Hallin:2021wme}, where the best 10 epochs based on the validation loss are used. 
    \item The \texttt{HistGradientBoostingClassifier}, which is a gradient boosted decision tree classifier implementation in \texttt{scikit-learn} \cite{Pedregosa:2011sk}. It histograms the classification features before training. We ensemble 50 independent trainings with different 50-50 training-validation splits as was done in Ref.~\cite{Finke:2023ltw}.
    \item The \texttt{AdaBoostingClassifier}, which is an adaptive boosted decision tree classifier implementation in \texttt{scikit-learn} \cite{Pedregosa:2011sk}. We build our implementation in analogy to Ref.~\cite{Finke:2023ltw} and perform an ensembling of 10 independently trained classifiers using random 50-50 splits of the training dataset.
\end{enumerate}
A more detailed explanation of these classifiers as well as their hyperparameters can be found in App.~\ref{app:hp settings}.

\subsection{Performance Metrics}
\label{subsec:performancemetric}

Throughout this work, we use the following performance metrics:
\begin{enumerate}
    \item The maximum SIC value with a statistics cut-off at a relative statistical error on the background efficiency of 20\%. This is a supervised metric for the anomaly detection performance calculated on the large held-out test set defined in Sec.~\ref{subsec:dataset}. We refer to this metric as ``max SIC" in the following.
    \item The BCE loss calculated on the validation set, which is the standard metric used to perform model optimization or select the best epochs during training (see e.g. Refs.~\cite{Hallin:2021wme, CMS:2024nsz}).
    \item The maximum ARGOS calculated on the validation set as defined in Eq.~(\ref{eq:ARGOS}). We refer to this metric as ``max ARGOS" in the following.
\end{enumerate}

\begin{figure*}[ht]
        \centering
        \includegraphics[width=\textwidth]{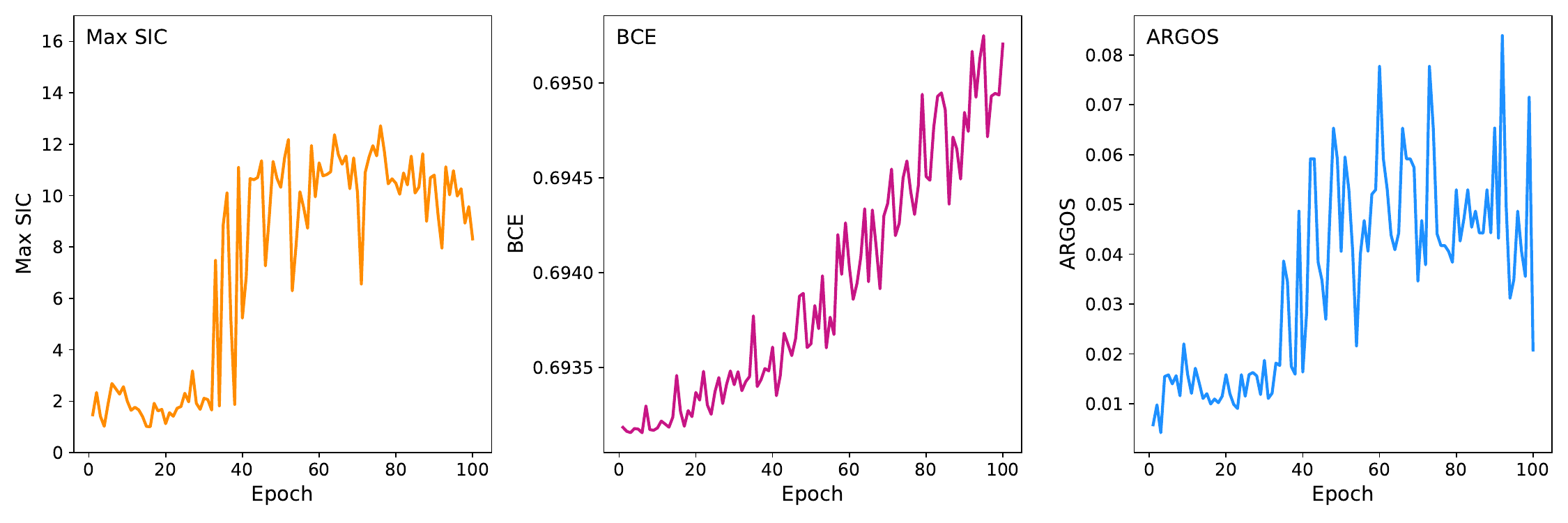}
        \caption{Example of metrics tracked throughout a NN training with $N_{sig}=400$ signal events using the default hyperparameters. We show the supervised max SIC metric (left) evaluated on the test set as well as BCE (middle) and max ARGOS (right), both evaluated on the validation set.}   
        \label{fig:IAD metrics NN}
\end{figure*}

\section{Experiments}\label{sec:tests}
There are a number of different applications for a metric that can analyze the performance of weakly supervised anomaly detection. Here, we discuss four examples, namely: selecting the best epochs to ensemble from a neural network training; optimizing hyperparameters; selecting an architecture; and selecting features for anomaly detection. However, we first highlight the correlation of ARGOS and SIC.

\subsection{Correlation of ARGOS and SIC}\label{sec:correlation}

In Fig.~\ref{fig:IAD metrics NN}, we show our three performance metrics (see Section~\ref{subsec:performancemetric}) tracked throughout a NN training for the IAD setup (see Sec.~\ref{subsec:IAD}). 

Comparing the standard BCE validation loss to the max SIC, there is no visible correlation. Instead, the validation loss very quickly shows overtraining, where the classifier tries to tell apart indistinguishable background events on the basis of statistical fluctuations. As the BCE loss takes the average over all events and most events are background, they dominate the metric. 

ARGOS instead focuses on the events with the highest anomaly score, since max ARGOS is found for large background rejection $1/\epsilon_{B}$. Therefore it is not as sensitive to background overtraining. While we expect an almost perfect correlation between max SIC and max ARGOS according to Eq.~(\ref{eq:ARGOSandSIC}), the validation-set ARGOS is severely statistics limited by the small number of signal events in the data. Thus the max ARGOS has a slightly noisy correlation with the max SIC but shows the same underlying trend. 

In this particular example, the signal is not found at the beginning of the training but instead after about 30 epochs, where the BCE already shows significant overtraining. This is not the case for every network initialization but it is common for the max SIC to show a delayed onset of the signal identification for lower signal injections, which the BCE is unable to identify. Hence, using early stopping could be counter productive.

\subsection{NN epoch selection}\label{sec:epoch selection}
\begin{figure*}[ht]
        \centering
        \includegraphics[width=\textwidth]{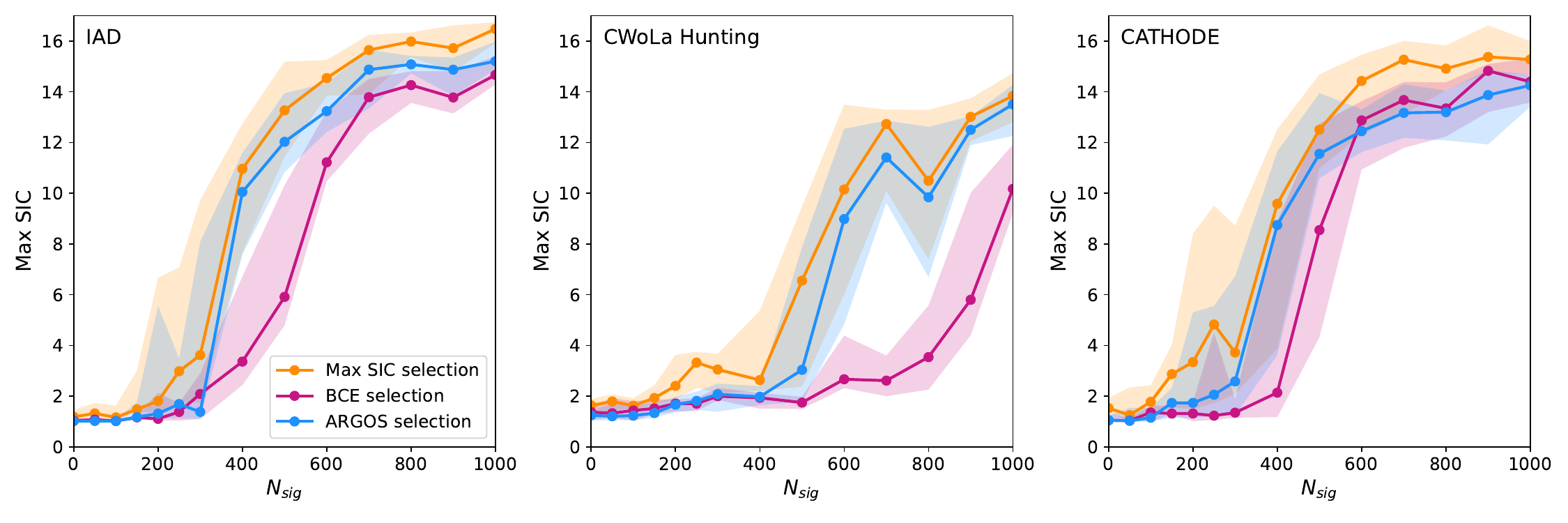}
        \caption{Anomaly detection performance (max SIC) as a function of the number of signal events $N_{sig}$ after epoch selection.  The epoch selection is performed using the supervised benchmark metric max SIC and the two data-driven metrics (max ARGOS and BCE), shown for IAD (left), CWoLa Hunting (middle) and CATHODE (right).}   
        \label{fig:Epoch selection NN}
\end{figure*}

As noted in Sec.~\ref{subsec:classifiers}, we ensemble the ten best epochs for the NN to obtain a lower variance and to stabilize the anomaly-detection performance. Which epochs are selected strongly depends on the metric used, as can be seen in Fig.~\ref{fig:IAD metrics NN} for example. The impact of ensembling based on our three tested metrics for the IAD, CWoLa Hunting and CATHODE is shown in Fig.~\ref{fig:Epoch selection NN}. We show the median anomaly detection performance of ten trainings as well as an error band representing the 16 to 84 percent quantiles.

\begin{figure*}[ht]
        \centering
        \includegraphics[width=0.66\textwidth]{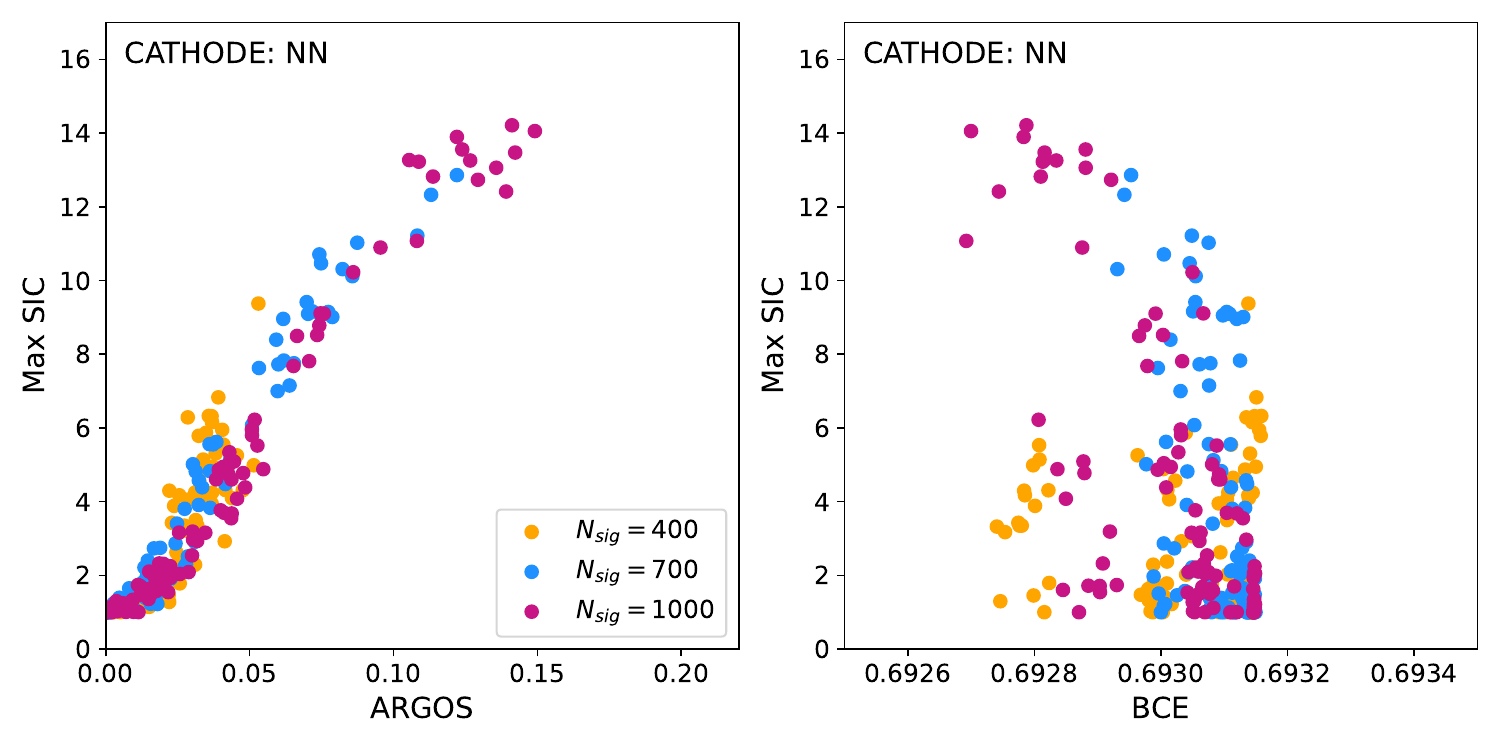}
        \caption{Correlation between median anomaly detection performance max SIC and two data-driven metrics max ARGOS (left) and BCE (right) for all 100 hyperparameter sets for CATHODE using the NN classifier at three example signal injections.}   
        \label{fig:HP metric correlation}
\end{figure*}

By definition, selecting the epochs based on our supervised target metric, the max SIC, results in the optimal performance in all cases. Selecting using the max ARGOS results in a slightly lower and somewhat noisier performance but clearly outperforms the selection based on the BCE loss, which especially falls short for low signal injections. 

This difference is particularly pronounced for CWoLa Hunting: when using BCE for epoch selection, the max SIC drops below 10 already at $N_{sig} = 900$, whereas with ARGOS this drop only occurs around $N_{sig} = 600$. In our setup, CWoLa Hunting generally has the worst quality background template of the three methods considered.\footnote{The background template of the IAD contains no mismodeling by definition and the flow-matching based density estimate used for CATHODE is highly accurate for the small feature space used here.} Therefore, the NN can learn to distinguish background template and signal region based on the background events only, which results in a lower validation BCE loss but does not transfer onto a better performance on the anomaly detection test set. Instead using the ARGOS metric, and in particular max ARGOS, allows the epoch selection to focus on the events with the highest anomaly score.

An interesting feature can be seen in the performance of the max SIC based epoch selection at $N_{sig}=0$, where the performance is slightly better than random. The fluctuations of the classifier weights in the absence of signal still lead to some epochs with non-zero anomaly-detection performance, which the max SIC identifies. Selecting the epochs using signal information can therefore enhance the false positive rate. Using ARGOS, which does not know anything about a potential signal, this issue is not present. We discuss why we do not generally expect an enhanced rate of false discoveries or enhanced background sculpting when selecting models on data in App.~\ref{app:sculpting}.

In the rest of the paper, we select epochs using the max ARGOS metric, as this is the best-performing data-driven metric for this application.

\subsection{Hyperparameter optimization}\label{sec:hp optimization}

\begin{figure*}[p]
        \centering
        \includegraphics[width=\textwidth]{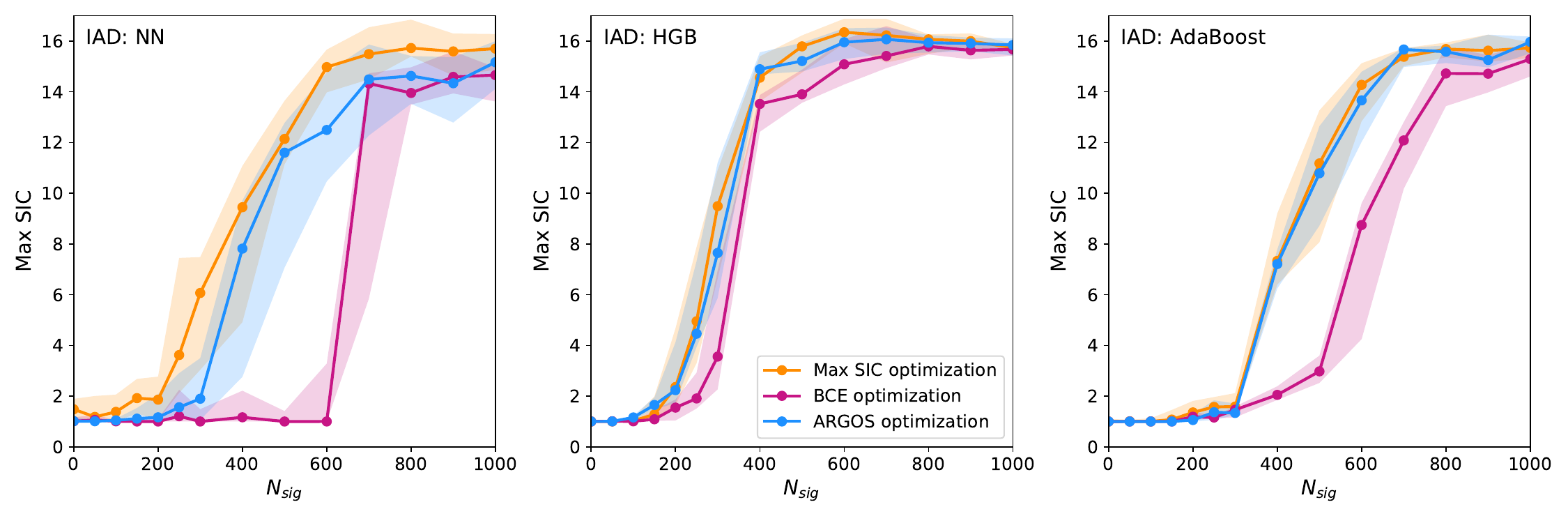}
        \includegraphics[width=\textwidth]{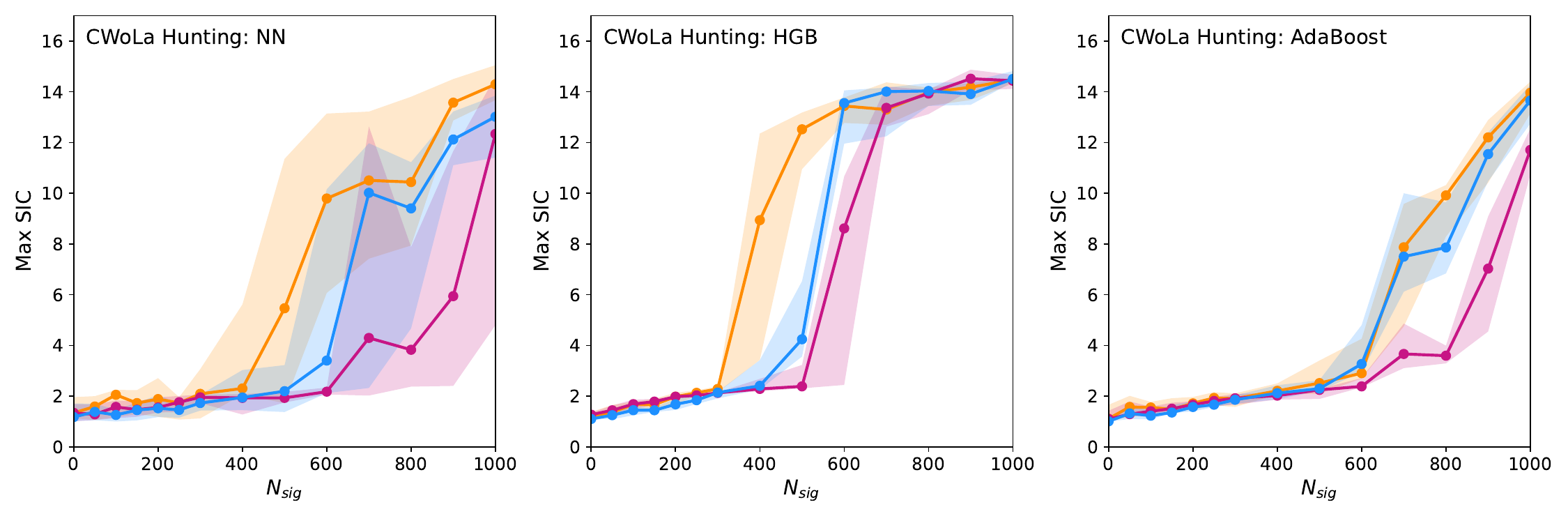}
        \includegraphics[width=\textwidth]{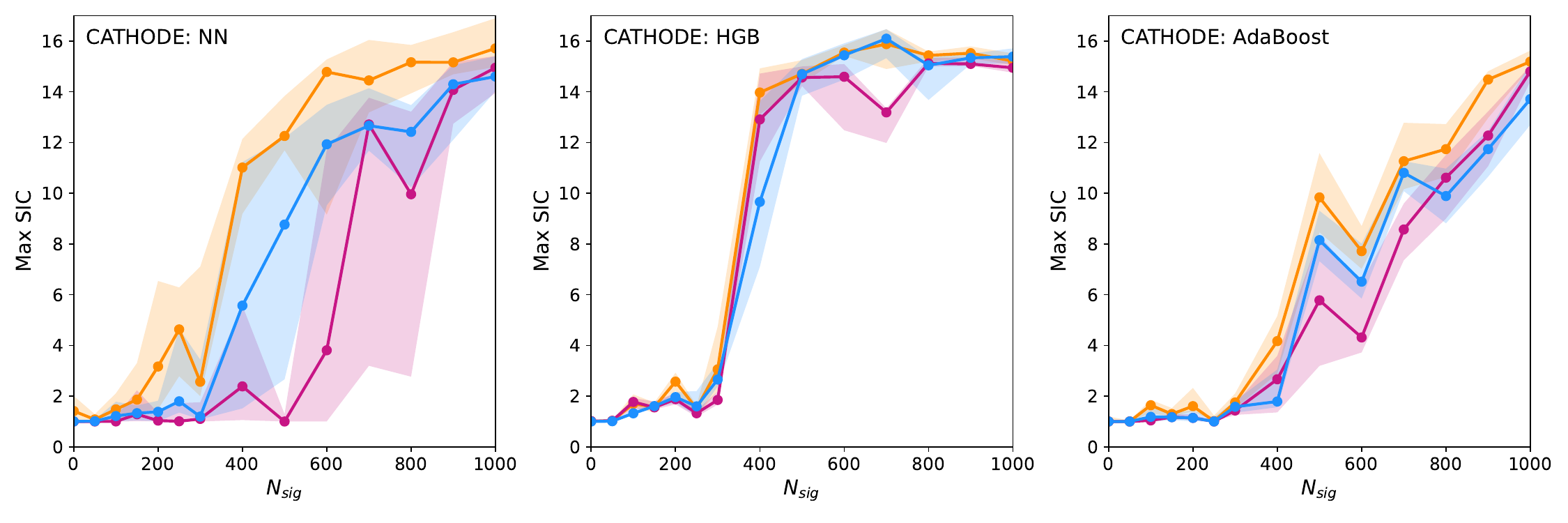}
        \caption{Anomaly detection performance (max SIC) after hyperparameter optimization for the NN (left), HGB (middle) and AdaBoost (right) classifiers. The optimization is performed using the supervised benchmark metric max SIC and the two data-driven metrics (max ARGOS and BCE), shown for IAD (top), CWoLa Hunting (middle), and CATHODE (bottom).}   
        \label{fig:HP optimization}
\end{figure*}

\begin{figure*}[ht]
        \centering
        \includegraphics[width=\textwidth]{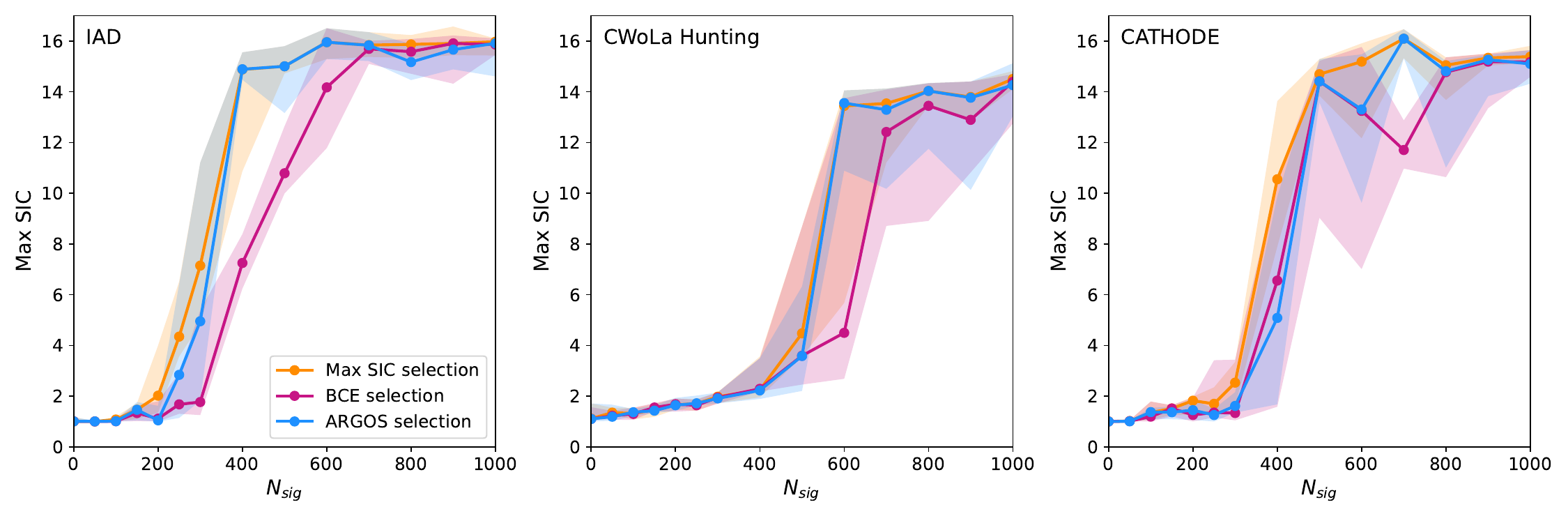}
        \caption{Anomaly detection performance (max SIC) after architecture selection based on the supervised metric max SIC as a benchmark and the two data-driven metrics max ARGOS and BCE for IAD (left), CWoLa Hunting (middle) and CATHODE (right).}   
        \label{fig:Model selection}
\end{figure*}

An especially important application of a data-driven metric for weak supervision is data-driven hyperparameter optimization. As an illustration, we perform this optimization using a random-search strategy, where we randomly sample different hyperparameters $N_{hp}=100$ times from a range of possible values as given in App.~\ref{app:hp settings}. Using each hyperparameter set, the classifier is trained and all three metrics are evaluated; BCE and ARGOS on the validation set and max SIC on the test set. Then the best hyperparameter set based on each metric is selected. This procedure is performed ten times for each signal injection $N_{sig}$ to obtain a stable median as well as 16 to 84 percent quantile error bands. A more detailed explanation of the procedure can be found in App.~\ref{app:hp optimization procedure}. 

In Fig.~\ref{fig:HP metric correlation}, we show the correlation between our data-driven metrics and the anomaly detection performance (max SIC) for the NN based CATHODE optimization at three different signal injections. The range of different hyperparameter configurations result in a large spectrum of performance.
We observe a much stronger correlation with max SIC for max ARGOS than for BCE, particularly for lower signal injections. 
For example, we  find relatively low BCE values for low max SIC values at $N_{sig}=400$. These low BCE values are likely achieved by optimizing on the differences between BT and SR background. We do not see a similar effect for the ARGOS metric. Analogous results are also obtained for the IAD and CWoLa Hunting and the other classifier architectures. Hence, we expect the optimization based on ARGOS to be more stable and to yield better results, especially at lower signal injections.

This is demonstrated in Fig.~\ref{fig:HP optimization}, which shows the significance improvement after hyperparameter optimization with the different metrics. Starting with the IAD results of the hyperparameter optimization in the top row of Fig.~\ref{fig:HP optimization}, we can once more see that an optimization using the supervised max SIC metric of course leads to ideal results. Second best in all cases is optimizing based on ARGOS, which in particular for the NN leads to a slightly noisier but still similar performance compared to optimizing on the max SIC. Optimizing on the BCE loss results in a worse anomaly detection performance, which is especially significant for the NN and AdaBoost. This effect is less prominent for the HGB, likely because its histogramming of the input data limits overtraining. These trends generally persist for CWoLa Hunting and CATHODE in the bottom two rows of Fig.~\ref{fig:HP optimization}, although the imperfect background templates cause the results to be slightly noisier, especially for CATHODE where the individual density estimation runs cause the BT quality to slightly vary between different signal injections.

\subsection{Architecture selection}\label{sec:architecture-selection}

In addition to optimizing the hyperparameters of a given architecture, one may also be interested in selecting the architecture itself. We train our three architectures (with hyperparameters optimized using ARGOS) on 50\% of our standard training and validation data. On the other half, we calculate BCE and ARGOS once the trainings are complete. We can then select the best architectures based on these two metrics as well as on the max SIC of the half-statistics training. The selected architecture is then trained on the full statistics and evaluated on the test set. We perform a separate architecture selection for each of the ten optimized hyperparameter sets obtained in Sec.~\ref{sec:hp optimization} at each signal injection. We plot these ten runs as a median and an error band corresponding to the 16 to 84 percent quantiles. A more detailed description of the architecture-selection procedure can be found in App.~\ref{app:architecture selection}.

Fig.~\ref{fig:Model selection} shows the results of the architecture selection for IAD, CWoLa Hunting and CATHODE. Again, selecting based on max SIC or ARGOS results in very similar performance, while a worse performance is observed when selecting based on the BCE. Interestingly, the effect is not the same for the IAD as it is for CATHODE or CWoLa Hunting. For the IAD, BCE results in a performance drop at higher signal injections compared to ARGOS and max SIC, whereas for CWoLa Hunting and CATHODE the dominant effect is a larger error band. This large error band originates from the same metric selecting different classifier architectures across the ten runs. For example, in the case of CATHODE,  the BCE metric largely chooses the HGB architecture, which leads to the median largely matching the optimal performance; but sometimes (three times out of ten) it chooses AdaBoost, which leads to the large downward shift of the error band. For CWoLa Hunting the selection of worse models becomes common enough to also lead to a downward shift of the median. 

In our case, it was necessary to perform multiple independent architecture selections in order to obtain error bands to help compare the different metrics. In an actual analysis, however, it may be more effective to select architectures based on the mean or median metric obtained from multiple trainings of the same hyperparameter configuration, which could further stabilize the performance and reduce downward fluctuations of the selection. 

\subsection{Feature selection}
\label{sec:feature selection}
\begin{figure}[t]
    \centering
    \includegraphics[width=\linewidth]{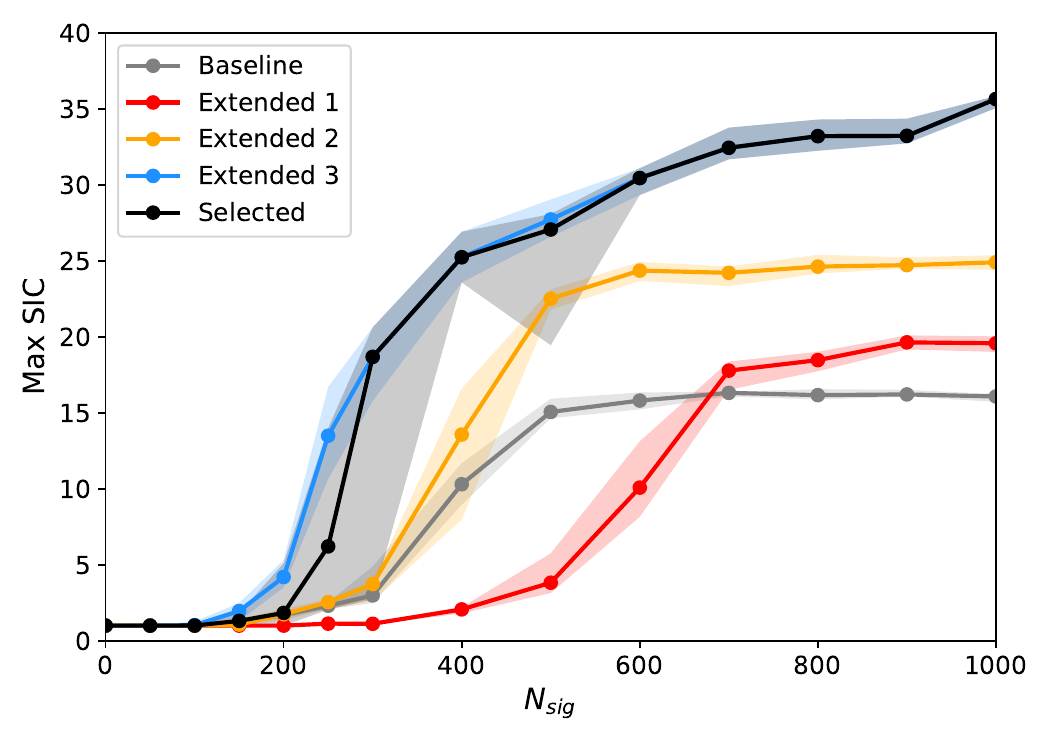}
    \caption{Anomaly detection performance max SIC for different subjettiness-based feature sets and for the feature set selection using max ARGOS in black. We use  the IAD setup with the HGB classifier. The baseline set is defined in Section~\ref{sec:setup}. The extended sets are introduced in Ref.~\cite{Finke:2023ltw}.}
    \label{fig:IAD feature selection}
\end{figure}

For anomaly detection to be truly signal model agnostic, feature sets should not be tuned on specific signals. To test the concept of choosing the best feature set based on data-driven metrics, we apply the procedure used for the architecture selection (see Sec.~\ref{sec:architecture-selection}) to the feature sets used in Ref.~\cite{Finke:2023ltw} for the IAD in Fig.~\ref{fig:IAD feature selection}. The largest feature set, extended set 3, is in this case most sensitive and at high signal injections always chosen by the ARGOS metric. At lower signal injections ARGOS sometimes also selects less performing feature sets as can be seen from the increased error band and the slightly earlier performance drop-off.

This IAD proof-of-concept study shows the potential viability of data-driven feature selection in general. 

\section{Conclusion and Outlook}
\label{sec:Conclusion}

In this paper, we have introduced the fully data-driven ARGOS metric for choosing the best anomaly detector. Given an accurate template of background events, we show that the ARGOS metric has a sound theoretical grounding -- it is monotonic with the significance improvement characteristic for any unknown signal in the data. This guarantees the effectiveness of the ARGOS metric for comparing arbitrary anomaly detectors.

We demonstrated the power of the ARGOS metric with a number of realistic use cases in the context of weak supervision. For two well-known methods (CWoLa Hunting and CATHODE), applications to epoch selection, hyperparameter optimization, and architecture selection showed that the ARGOS metric is a better proxy for the true anomaly detection performance than the binary cross-entropy loss, which is the default standard currently used in classifier-based weakly-supervised anomaly detection.

In this work, we employ the maximum of the ARGOS metric as a function of $\epsilon_\text{BT}$, i.e., we optimize the working point $\epsilon_\text{BT}$ and the anomaly detector at the same time. One could just as easily evaluate ARGOS at a specific background template efficiency $\epsilon_{\text{BT}}$, if $\epsilon_{\text{BT}}$ is fixed in a given analysis.

We stress that this is not the only context where the ARGOS metric can be used. For example, it is directly applicable to ANODE \cite{Nachman:2020lpy} and R-ANODE \cite{Das:2023bcj}, which are density-estimator-based resonant anomaly detection methods. Beyond resonant anomaly detection, different outlier detectors (e.g.\ autoencoders) could also be compared to one another if a background template is available. In contrast to the binary cross-entropy loss for classifier-based anomaly detection, ARGOS does not assume predictions to represent a class probability. ARGOS is a cut-based metric, which is only affected by the relative ordering of different events and invariant to rescaling or shifting of the entire anomaly score. Therefore, ARGOS can be applied to any anomaly score.

As an illustration of the potential for data-driven feature selection, we carried out a proof-of-concept study using the IAD setup, where ARGOS reliably identifies the best-performing feature sets at high signal injections. However, for a realistic background template, ARGOS will exhibit a bias towards worse background templates. The mitigation of this issue needs to be studied in future work in order for data-driven feature selection to be viable on real data.

\section*{Acknowledgements}

The authors thank Tobias Quadfasel for his contributions in the early stages of this project.
MH is supported by the Deutsche Forschungsgemeinschaft (DFG, German Research Foundation) under grant 400140256 - GRK 2497: The physics of the heaviest particles at the Large Hadron Collider. The research of MK and AM is supported by the DFG under grant 396021762 - TRR 257: Particle Physics Phenomenology after the Higgs Discovery. GK and LM acknowledge support by the DFG under Germany’s Excellence Strategy 390833306 – EXC 2121: Quantum Universe. DS is supported by DOE grant DOE-SC0010008.
Computations were performed with computing resources granted by RWTH Aachen University under project rwth0934.

\section*{Code}
The code for this paper can be found in Ref.~\cite{modelselectiongithub}.

\appendix

\section{Architectures and training}
\subsection{Implementation of the CATHODE Density Estimation}
\label{app:implementation}

In order to obtain state of the art results, we use a conditional flow matching (CFM) density estimator based on Ref.~\cite{Das:2024fwo}. In this implementation \cite{density_estimation}, the conditional vector field needed for CFM is learned using a \texttt{ResNet}-style~\cite{resnet} architecture from \texttt{nflows}~\cite{nflows}. The hyperparameters from Ref.~\cite{Das:2024fwo} are used for both the model and the training. For each signal injection, ten density estimators are trained and ensembled by combining their samples in order to limit fluctuations of the density estimation quality between the different signal injections.  

\subsection{Classifier implementations and hyperparameter settings}\label{app:hp settings}

\subsubsection{NN}

For the NN, we use a standard MLP based on the architecture used in Ref.~\cite{Hallin:2021wme} with an implementation in \texttt{PyTorch}~\cite{Ansel_PyTorch_2_Faster_2024} based on Ref.~\cite{sk-cathode}. We choose a fixed number of three hidden layers with 64 nodes each and ReLU activation. We train on the binary cross-entropy loss using the ADAM optimizer \cite{Kingma:2014ad} for 100 epochs. Other hyperparameters such as batch size, dropout, learning rate and other ADAM settings are optimized later on. Their default values can be found in Tab.~\ref{tab:hp_NN}. For the NN, we draw hyperparameters from a set of choices, which are also included in the table. 

We choose a 50-50 training and validation split, calculating BCE and max ARGOS on the validation set and max SIC on the test set after every epoch. After training, we ensemble the best 10 epochs as decided by the different metrics. This follows the ensembling strategy used in Ref.~\cite{Hallin:2021wme}, where the best 10 epochs based on the validation loss are used. The impact of choosing different metrics can be seen in Sec.~\ref{sec:epoch selection}. As a default, we pick the epochs on max ARGOS.

\begin{table}[t] 
    \def\arraystretch{1.5}
    \centering
    \begin{tabular}{l|l|l}   
    Hyperparameter      & Default & Choices \\\hline
     Learning rate      & $10^{-3}$ & $[0.01, 0.005, 0.001, 0.0005, 0.0001]$\\\hline
     Batchsize          & 128 & $[64, 128, 256, 512, 1024, 2048, 5096]$\\\hline
     Dropout            & 0   & $[0,0.1, 0.2, 0.3, 0.4, 0.5]$\\\hline
     Weight Decay       & 0   & $[0,10^{-5},10^{-4},10^{-3},10^{-2}]$\\\hline
     Momentum           & 0.9 & $[0.8, 0.9, 0.95, 0.99]$\\\hline
    \end{tabular}
    \caption{Default NN hyperparameter settings and discrete choices sampled from for hyperparameter optimization}
    \label{tab:hp_NN}
\end{table}

\subsubsection{HGB}

\begin{table}[t]
    \centering  
    \def\arraystretch{1.5}
    \begin{tabular}{l|l|l|l|l} 
    Hyperparameter & Default & Sampling & Min & Max \\\hline
     Learning rate      & 0.1 &Log10Uniform & $10^{-4}$ & $10^{-0.1}$ \\\hline
     Max. Iterations    & 200 & UniformInt        & 2 & 200 \\\hline
     Max. Leaf Nodes    & 31 & UniformInt        & 2 & 100 \\\hline
     Max. Depth         & None & UniformInt        & 2 & 15 \\\hline
     Max. Bins          & 255 & UniformInt        & 31 & 255 \\\hline
     L2 Regularization  & 0 & Uniform           & 0 & 10 \\\hline
    \end{tabular}
    \caption{Default HGB hyperparameter settings and distributions sampled from for hyperparameter optimization}
    \label{tab:hp_HGB}
\end{table}

\texttt{HistGradientBoostingClassifier} is a gradient boosted decision tree classifier implementation in \texttt{scikit-learn} \cite{Pedregosa:2011sk}, which histograms the classification features before training. It is currently state of the art for weakly supervised anomaly detection \cite{Finke:2023ltw, Freytsis:2023cjr} with high-level feature. Following Ref.~\cite{Finke:2023ltw}, we perform 50 independent trainings with different 50-50 training-validation splits, which we ensemble by taking the mean over the predictions. Our default hyperparameters are largely equivalent to the default of the \texttt{scikit-learn} package and can be found in Tab.~\ref{tab:hp_HGB}. For hyperparameters optimization, we sample hyperparameters from the distributions also shown in the table.

\begin{figure*}[ht]
        \centering
        \includegraphics[width=\textwidth]{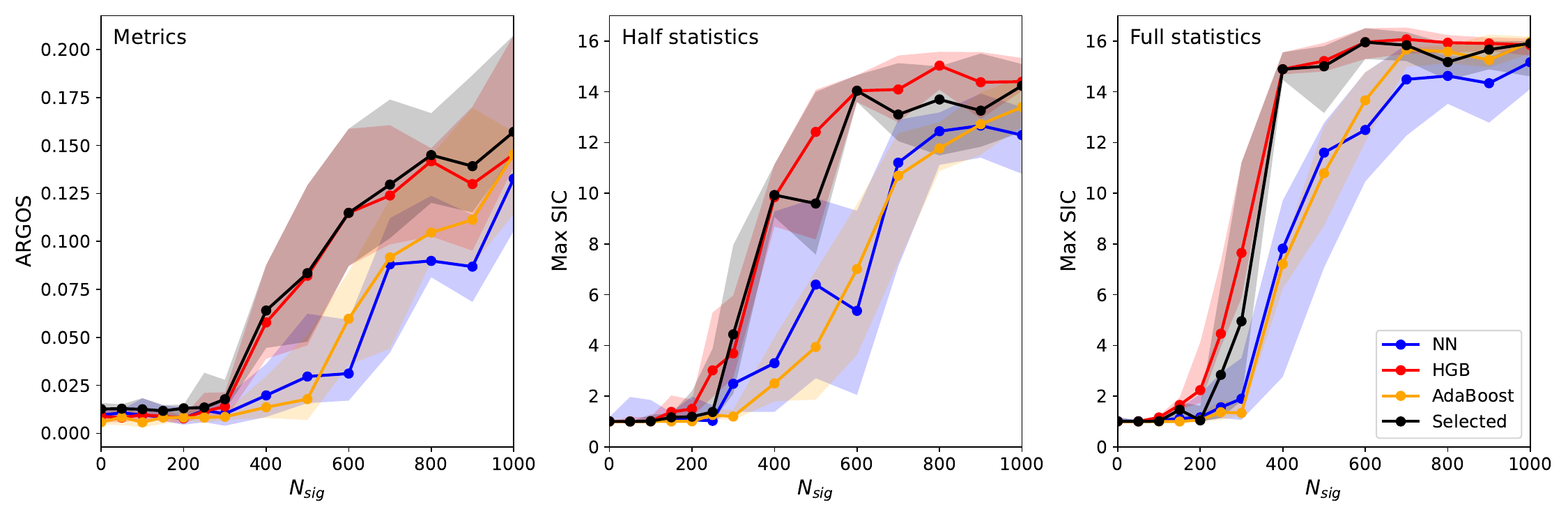}
        \caption{Architecture selection for IAD using max ARGOS. Left: Max ARGOS calculated on half the data when trained on the other half. Middle: Max SIC of the trainings of half the data. Right: Max SIC of full statistics training. In each case the selected runs are shown in black. Note that the x-axis shows the full statistics signal injection values even for the half statistics runs.}   
        \label{fig:model selection IAD ARGOS}
\end{figure*}

\subsubsection{AdaBoost}

\texttt{AdaBoostingClassifier} is an adaptive boosted decision-tree classifier implementation in \texttt{scikit-learn}, which was investigated as an alternative to HGB in \cite{Finke:2023ltw} but did not achieve the performance or robustness of HGB. We build our implementation in analogy to \cite{Finke:2023ltw} and perform an ensembling of 10 independently trained classifiers using random 50-50 splits of the training dataset. AdaBoost does not make use of a validation set. However, as training is deterministic, the random data split results in variations between different trainings, which can be used to obtain a better performance through ensembling. Our default hyperparameters reflect the default used for AdaBoost in \cite{Finke:2023ltw} and can be found in Tab.~\ref{tab:hp_AdaBoost}. The distributions used to randomly sample hyperparameters for optimization can be found there, as well.

\begin{table}[t]
    \centering  
    \def\arraystretch{1.5}
    \begin{tabular}{l|l|l|l|l}   
    Hyperparameter & Default & Sampling & Min & Max \\\hline
     Learning rate     & 1 & Log10Uniform & $10^{-4}$ & $10^{-0.1}$ \\\hline
     Number Estimators  & 50 & UniformInt        & 2 & 100 \\\hline
     Max. Leaf Nodes    & 31 & UniformInt        & 2 & 100 \\\hline
     Max. Depth         & None & UniformInt        & 2 & 15 \\\hline
     Max. Samples per Leaf & 20 & UniformInt     & 1 & 100 \\\hline
    \end{tabular}
    \caption{Default AdaBoost hyperparameter settings and distributions sampled from for hyperparameter optimization}
    \label{tab:hp_AdaBoost}
\end{table}

\section{Hyperparameter optimization procedure}
\label{app:hp optimization procedure}

As explained in Sec.~\ref{sec:hp optimization}, the hyperparameter optimization in this work follows a random search strategy with a random sampling of $N_{hp}=100$ hyperparameter configurations. The distributions these hyperparameters are sampled from can be found in App.~\ref{app:hp settings}. A classifier is trained with each of these configurations, BCE and max ARGOS are evaluated on the validation set and max SIC on the test set. We use slightly different strategies of stabilising the metrics and speeding up the computation time for each of our architectures: 
\begin{itemize}
    \item For the NN, each network is trained for 20 epochs, with the epoch number set to 100 after optimization. All three metrics are evaluated after each epoch and the best value for each (minimum for BCE, maximum for max SIC and max ARGOS) is selected. 
    \item For AdaBoost and HGB, we do not ensemble during the optimization, instead the metrics evaluated after training are averaged over ten runs to obtain more stable results. 
\end{itemize}

Based on the metric values obtained in this way, the best hyperparameter configuration based on each metric is selected. This entire procedure is performed ten times for each signal injection $N_{sig}$.

\section{Details of the architecture selection}
\label{app:architecture selection}

In this appendix, we explain the procedure of the architecture selection in more detail using the example of picking the best architecture for the IAD using ARGOS in Fig.~\ref{fig:model selection IAD ARGOS}. In Fig.~\ref{fig:Model selection} of Sec.~\ref{sec:architecture-selection} this is shown as a single line with an error band but in Fig.~\ref{fig:model selection IAD ARGOS} the entire procedure can be seen.

\begin{figure*}[ht]
        \centering
        \includegraphics[width=\textwidth]{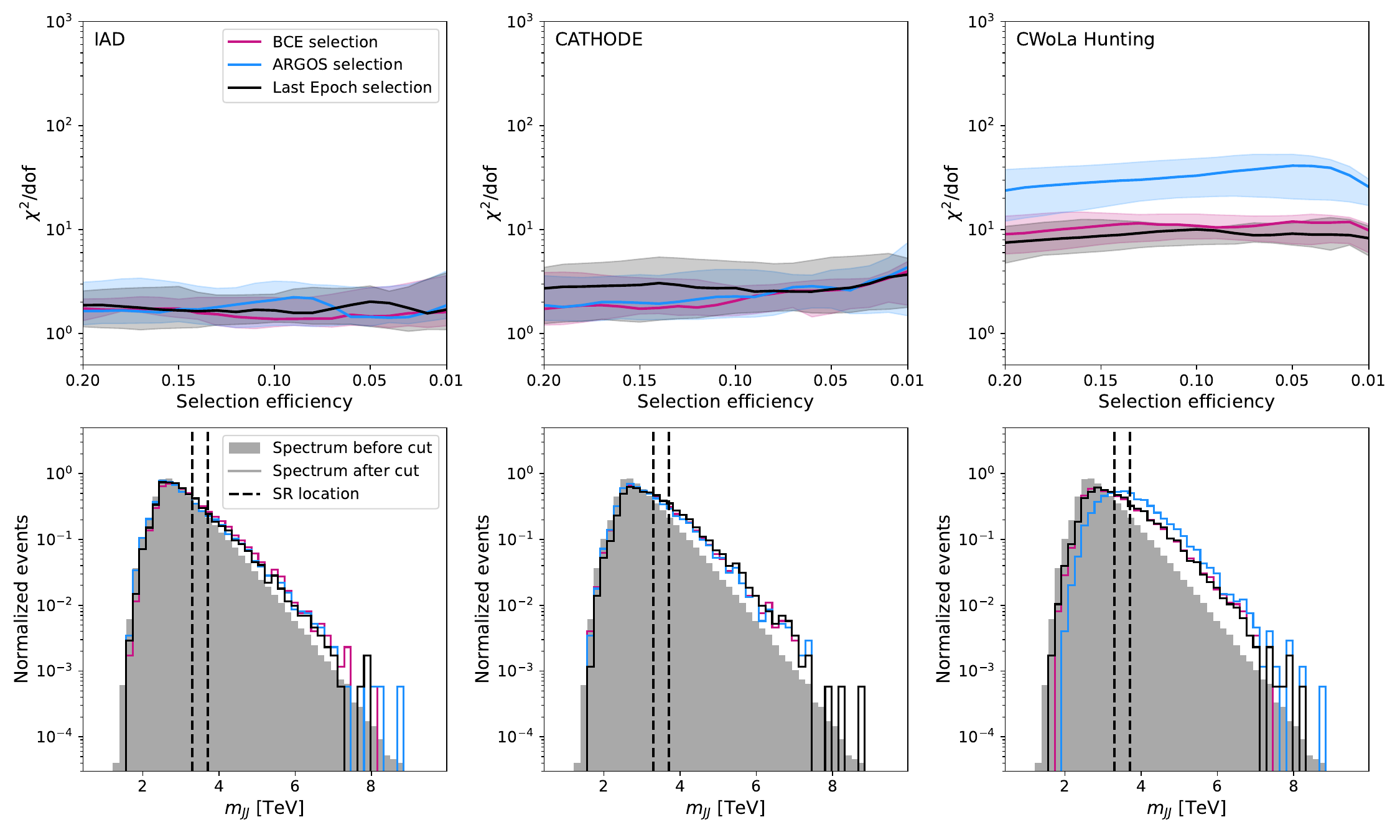}
        \caption{Background sculpting for IAD (left), CATHODE (middle) and CWoLa Hunting (right) when selecting epochs based on validation BCE loss, max ARGOS or selecting the last epoch characterised by the $\chi^2$ metric from \cite{Hallin:2022eoq} (top) and example spectra obtained when selecting $1\%$ most anomalous events (bottom).}   
        \label{fig:background sculpting}
\end{figure*}

The first step of our architecture selection is to train our architecture options -- NN, HGB and AdaBoost -- on 50\% of our normal training and validation set. The anomaly detection performance characterized by the max SIC of these runs is shown in the middle panel of Fig.~\ref{fig:model selection IAD ARGOS}. On the other half of the data, we then calculate the ARGOS metric, the result of which is shown in the left panel of Fig.~\ref{fig:model selection IAD ARGOS}. For both of these panels we show the median and 16 to 84 percent quantiles of ten runs as lines and error bands respectively. These ten runs contain not only ten independent trainings but also ten different hyperparameter configurations per model as we use the results from Sec.~\ref{sec:hp optimization}. For each of these sets of ten runs, we then pick the best-performing architecture between NN, HGB and AdaBoost based on the obtained max ARGOS values and plot the resulting set of ten again as median and quantile-error band. We perform this 50-50 split of the data set in order to have the metric values used for architecture selection be calculated on a held-out evaluation set. As all classifiers use the validation set in one form or another, this is not the case when directly selecting on the validation set. 

For this set of selected architectures, we then run another training using the full statistics, which leads to the black line and error band in the right panel of Fig.~\ref{fig:model selection IAD ARGOS}. For comparison, we also show the full statistics performance of the different architectures. One can see that the final performance of the selected runs largely matches that of the best performing architecture, namely the HGB, with only some downward fluctuations. 

\section{False discoveries and background sculpting}
\label{app:sculpting}

When optimizing on data, it is important to understand the effect that this has on the statistics of an analysis. In our case, we seek to understand whether an increased rate of false discoveries may follow, which could stem from one of two issues:

\begin{enumerate}
    \item Selecting the best classifier on data could lead to ``p-hacking":
    The selected classifier may prioritize purely statistical fluctuations in the training data, leading to a look elsewhere effect \cite{Hein:2025ysv} and miscalibrated significances. 
    \item The selected classifier may prioritize regions of the data with the most systematic background mismodeling, which could lead to false excesses in both fit-based bump hunts (as performed in Refs.~\cite{ATLAS:2020iwa,CMS:2024nsz,ATLAS:2025obc,Gambhir:2025afb}) and in analyses with direct background estimation \cite{Das:2024eie}. 
\end{enumerate}

The first issue can be completely eliminated by evaluating on an independent test set. This will be the case for the remainder of this appendix. 

Regarding the second issue,
the classifier most able to distinguish between data and background is fundamentally the better anomaly detector. If this leads to false positives due to background mismodeling, this is not a fault of the anomaly detector per se.

Nevertheless, we will investigate this issue in more detail, to see what the effect of improved model selection has on it. 

Shown in Fig.~\ref{fig:background sculpting} is the $\chi^2$ histogram difference for the background $m_{JJ}$ distribution before and after a cut on the anomaly score, as a function of background selection efficiency. This metric was previously introduced in Ref.~\cite{Hallin:2022eoq} and is a way to quantify the effect of background sculpting due to a cut on the anomaly score. We compare the
sculpting for an anomaly detector selected using the BCE loss (as used in Ref.~\cite{CMS:2024nsz}), the ARGOS metric and using the last epoch. We see  that for the IAD and CATHODE the sculpting does not change significantly based on the different model selection methods. 

For CWoLa Hunting, the amount of sculpting is already more severe compared with IAD or CATHODE when model selecting with BCE loss or using the last epoch. This is further exacerbated when the selection is performed with ARGOS. This is entirely expected as the background in the SR is not perfectly modeled by the SB data, and with the ARGOS metric we are picking this up more effectively.

If significances are obtained through direct background estimation the magnitude of the systematic shift between BT and SR needs to be estimated in any case for the given classifier setup \cite{Das:2024eie}. Using a better classifier based on ARGOS optimization, the systematic shift might increase. However, estimating the systematic shift properly, an increased rate of false discoveries will be avoided.

\bibliographystyle{apsrev4-1}
\bibliography{HEPML, other}
\end{document}